\documentclass[twocolumn]{aastex631}

\newcommand\latex{La\TeX}

\usepackage[caption = false]{subfig}
\usepackage{comment}
\usepackage{lineno}

\begin{document}

\title{A second candidate magnetic helium core white dwarf and 3 other variable white dwarfs in the globular cluster NGC~6397}

\correspondingauthor{Manuel Pichardo Marcano}
\email{mmarcano@fisk.edu}

\author[0000-0003-4436-831X]{Manuel Pichardo Marcano}
\affiliation{Department of Life and Physical Sciences, Fisk University, 1000 17th Avenue N., Nashville, TN 37208, USA}
\affiliation{Department of Physics and Astronomy, Vanderbilt University, 6301 Stevenson Center Lane, Nashville, TN 37235, USA}

\author[0000-0002-9396-7215]{Liliana E. Rivera Sandoval }
\affiliation{Department of Physics and Astronomy, University of Texas Rio Grande Valley, Brownsville, TX
78520, USA }

\author[0000-0003-0976-4755]{Thomas J.  Maccarone  }
\affiliation{Department of Physics and Astronomy, Texas Tech University, Lubbock, TX 79409, USA}

\author[0000-0001-7209-3574]{Rene D. Rohrmann}
\affiliation{Instituto de Ciencias Astronómicas, de la Tierra y del Espacio (CONICET-UNSJ), Av. España 1512 (sur), 5400 San Juan, Argentina}

\author[0000-0003-2771-7805]{Leandro G. Althaus}
\affiliation{Grupo de Evoluci\'on Estelar y Pulsaciones. Facultad de Ciencias Astronómicas y Geofísicas,Universidad Nacional de La Plata, Paseo del Bosque s/n, 1900 La Plata, Argentina}
\affiliation{IALP - CONICET, La Plata, Argentina}

\author[0000-0003-3944-6109]{Craig O. Heinke}
\affiliation{Department of Physics, University of Alberta, CCIS 4-183, Edmonton, AB, T6G 2E1, Canada}

\author[0000-0003-1535-0866]{Diogo Belloni}
\affiliation{Departamento de F\'isica, Universidad T\'ecnica Federico Santa Mar\'ia, Av. España 1680, Valpara\'iso, Chile}

\author[0000-0003-2506-6041]{Arash Bahramian}
\affiliation{International center for Radio Astronomy Research Curtin University, GPO Box U1987, Perth, WA 6845, Australia}



\begin{abstract}

Using archival Hubble Space Telescope observations, we report the discovery of four variable low-mass white dwarfs ($0.18 \, M_\odot \leq M \leq 0.5 \,M_\odot$) in the globular cluster NGC 6397. One source exhibits a periodic optical modulation of  $5.21 \pm 0.02$ hours, which we interpret as potentially due to the rotation of a magnetic helium core WD (He WD). This makes this candidate the second magnetic He WD in NGC~6397, and one of the few He WDs with a measured rotation period. The other three candidates show aperiodic variability, with a change in magnitude ranging from $\sim 0.11-0.6$. These discoveries highlight the importance of high-cadence photometric surveys in dense stellar environments. Follow-up spectroscopic observations are needed to confirm the nature of the variability of these systems.

\end{abstract}



\section{Introduction} \label{sec:intro}

NGC~6397 is the closest core-collapsed globular cluster. Due to its proximity (2.482 $\pm 0.019$ kpc, \citealt{BaumgardtDistance2021}) and low reddening ($E(B - V) = 0.18$, \citealt{hansen2013extintion}), it has been extensively observed in the optical and the ultraviolet (UV) with the {\it Hubble Space Telescope} (HST). One population that has been studied in some detail is the white dwarf (WD) population. \cite{HansenWDSequence2007ApJ...671..380H} used deep HST observations to measure the age of this globular cluster from the Carbon-Oxygen WD (CO WD) cooling sequence. Later, \cite{Richer2008AJ....135.2141R}  imaged a single field $5'$ southeast of the cluster center with HST/Advanced Camera for Surveys (ACS) for 126 orbits targeting the CO WD sequence. Simultaneous exposures taken with the Wide Field and Planetary Camera 2 (WFPC2) imaged the core of NGC~6397. More recently, \cite{Torres2015WDsNGC6397} used a population synthesis to study the WD population and estimate its age, and \cite{Vitral2022MNRAS.514..806V} using Gaia and HST observations, concluded that the inner mass of NGC~6397 should be dominated by hundreds of massive WDs as predicted by \cite{Kremer2021ApJ...917...28K}.

Another population of WDs that is predicted to exist in globular clusters are extremely low-mass  (ELM) helium core WDs (He WD). These are the remnants of the unignited helium cores of evolved stars that were stripped of their envelopes, most probably due to mass transfer in close binaries \citep[e.g.][]{Marsh1995MNRAS.275..828M}, or in some high metallicity stars maybe due to strong winds in the red giant phase \citep{KilicSingleELM2007ApJ...671..761K},a model that is highly unlikely to apply in low metallicity globular clusters like NGC~6397. As the product of binary evolution, they are important for understanding the formation history of compact binaries in globular clusters and their 
study 
could allow for better characterization of the gravitational wave foreground for the upcoming Laser Interferometer Space Antenna (LISA) mission \citep{AmaroLISA2023}.

The number of He WDs in a globular cluster is expected to depend on the primordial binary fraction \citep{KremerWDs2021ApJ...917...28K}, 
as these systems are thought to be products of binary evolution. In NGC~6397, \cite{Cool98} reported faint and hot stars that did not show the characteristic stochastic broad-band variability on timescales of
minutes to hours, also called "flickering", that is characteristic of cataclysmic variables. While the origin of this variability is not well understood most scenarios link flickering to accretion of matter \citep[e.g.][]{Bruch(1992)Flickering}. The UV bright stars that did not show flickering were denoted as “non-flickerers” (NFs). One of them, NF1 was later spectroscopically confirmed by \cite{edmonds_cataclysmic_1999} as a He WD with a low mass of $\sim  0.25 M_\odot$. Other He WD candidates in NGC~6397 were found via photometry and their position on the color-magnitude diagram (CMD), i.e. parallel to, but brighter than, the CO WD sequence. Using HST photometry, \cite{taylor_helium_2001} reported 6 candidate He WDs and showed that the radial distribution for He WDs is significantly more centrally concentrated than the main-sequence stars, suggesting that 
these objects 
are in binary systems. 
\cite{Strickler2009ApJ...699...40S} raised the total to 24 good candidate He WDs in NGC~6397 
based on their position in the CMDs. \cite{Strickler2009ApJ...699...40S} determined photometric masses in the range of $0.2-0.3 M_\odot$, and based on the locations of the objects within the cluster, determined dynamical masses similar to those of the blue stragglers ($\sim 0.8-2 M_\odot$). The difference in the masses argues that these must typically 
have 
companions that are rather heavy for globular cluster objects; given that the colors of these objects are not affected by their companions, the companions are 
other compact objects. 
These 
systems 
are likely dominated by 
double WD binaries with a He WD and a dark CO WD \citep{HansenWDSequence2007ApJ...671..380H}.
He WDs also have been found as companions to blue stragglers \citep[e.g.][]{Geller2011Natur.478..356G} and as companions to neutron stars (NSs), in either ultra-compact X-ray binaries  \citep[e.g.][]{Stella1987} or detached millisecond pulsar binaries \citep[e.g.][]{vanKerkwijk2010ApJ...715...51V,RiveraSandoval2015MNRAS.453.2707R}.

Extremely low-mass (ELM) WDs have also been extensively studied in the Galactic field, in great part by the ELM survey series both in the northern sky \citep{ELMBrown2010}, and in the southern sky \citep{ELMSKosakowski2023ApJ...950..141K}. Their variability was studied by \cite{BellELMVars2017ApJ...835..180B} and others using ground-based data and space-based observations \citep{LopezPulsatinELMTESS2021ApJ...922..220L}. ELM photometric variability can have many causes, 
most of which involve 
their binarity (in the Galactic field their binary fraction is almost 100\%; \citealt{BrownBinaryELM2011ApJ...730...67B,Kilic2011ApJ...727....3K}); eclipses, ellipsoidal variations, Doppler beaming or pulsations \citep{HermesELMPulsation2013ApJ...765..102H}.

In globular clusters, the variability and binarity of He WDs have been poorly studied, partly due to the lack of high cadence data and partly due to the difficulty of doing photometry in crowded fields. This is 
worse in the centers of globular clusters, where we need angular resolution that can be achieved from the ground only via adaptive optics and from space only with HST and JWST. One new variable He WD candidate was reported by \cite{PichardoMarcano2023} with a periodicity of 18.5 hours, which we attributed to the rotation period of the WD, with the modulation driven by magnetism in the He WD. This has been found before in CO WDs in the Galactic field, but little is known about the magnetic fields of He WDs. 

Another population related to He WDs that shows variability and hints of magnetism in globular clusters is hot subdwarfs. These are core helium-burning stars \citep{Heber2009,Heber2016PASP..128h2001H} and are also thought to be formed via binary evolution \citep{2002MNRAS.336..449H,PelisoliBinarysd2020A&A...642A.180P,2024MNRAS.52711184A} and will eventually evolve to become low-mass hybrid WDs \citep[$\lesssim 0.5 M_\odot$, e.g.,][]{Ibsen1985ApJS...58..661I,Han2000MNRAS.319..215H}, i.e., the resulting CO WDs will be hybrid, rich in helium ($\gtrsim $ several~per~cent), as they only burn helium partially, in contrast with more massive WDs of this type which are deficient in helium (${\lesssim1}$~per~cent). By studying their light curves in globular clusters, \cite{Momany2020MagneticHotNatAs...4.1092M} reported the detection of photometric variability consistent with spots that could be due to a magnetic field. In the field, there are some magnetic candidates identified based on their variability \citep{Jeffery2013MNRAS.429.3207J,Geier2015A&A...577A..26G,Balona2019MNRAS.485.3457B, PelisoliMagsd2022MNRAS.515.2496P}. These populations then offer a great opportunity to study the origin of magnetism in low-mass WDs.

In this work we analyzed archival data from HST of 4 candidate He WDs based on their position on the color-magnitude diagram. We report one periodic source and 3 other aperiodic variable sources. 

\section{Data Analysis and Results}

\subsection{Data}

The objects presented in this work were found as variables, and to the left of the main-sequence (MS), as part of the Survey for Compact Objects and Variable Stars (SCOVaS) \citep{ScovasAAS}. SCOVAS is a survey for cataclysmic variables and compact binaries in Galactic globular clusters using multi-wavelength archival data from HST.

For NGC 6397, we used  WFPC2 data from the parallel field that observed the core of the cluster from the HST large program GO-10424 (PI H. Richer, with results published in \citealt{Richer2006Sci...313..936R}). The original data set consists of 126 orbits using the Wide Field and Planetary Camera 2 (WFPC2). Each orbit is divided into three exposures in three different filters (F814W, F606W, and F336W). For this work, we used the 126 individual exposures in the filter F336W with exposure times ranging from 500-700 seconds taken in 
2005 between mid-March and April (2005-03-13 to 2005-04-08). The minimum separation between consecutive data points is 74 minutes, the maximum separation between consecutive data points is 3.2 days, and the total baseline is 26 days.

We also used the GO-10257 dataset (PI: Anderson, with results published in  \citealt{cohn_identification_2010}), which provides ACS, Wide Field Channel (ACS/WFC) imaging of the central region of NGC~6397 in the filters F435W (B), F625W (R), and F658N ($H\alpha$). 

For the photometry of both datasets, we used the DOLPHOT software package  \citep{2000PASP..112.1383D}. For  WFPC2, we supplied DOLPHOT with the calibrated single-exposure image 
(c0m) WFPC2 files and the drizzled .drz image as the reference frame for alignment, and for the ACS/WFC we supplied the CTE-corrected 
.flc images and a .drz image as 
the reference frame.  
The final output for both runs from the software lists the position of each star relative to the reference image, together with the measured aperture-corrected magnitudes calibrated to the Vega system for the individual exposures, along with some diagnostic values. We limited the data to measurements containing an error flag of zero, meaning that the star was recovered extremely well in the image without contamination due to cosmic rays.

We also made use of the \emph{HST} UV Globular Cluster Survey (HUGS; \citealt{Piotto2015AJ,Nardiello2018MNRAS,HUGSdoi}) and the Hubble Space Telescope Atlases of Cluster Kinematics (HACKS;  \citealt{HACKSLibralato2022,HACKSdoi}). The HUGS catalog\footnote{https://archive.stsci.edu/prepds/hugs/} provides photometry in 5 different bands (F275W, F336W, F438W, F606W, and F814W). The HACKS dataset also provides a photometric and  proper motion catalogue for NGC~6397\footnote{http://dx.doi.org/10.17909/jpfd-2m08}. We used both public catalogues to build CMDs of the cluster to confirm that our photometric results are consistent with previous studies, and to confirm the membership of our sources. 

As reported in \cite{Nardiello2018MNRAS}, the membership probabilities in the HUGS catalog were computed using
the local-sample method similar to the work of \cite{Libralato2014A&A...563A..80L,Bellini2009A&A...493..959B,BalaguerPM1998A&AS..133..387B}. 

\subsection{Selecting Variables }

We identify four sources as candidate periodic variables based on the False Alarm Probability (FAP) of the Lomb-Scargle periodogram \citep{Lomb1976Ap&SS..39..447L,Scargle1982ApJ...263..835S}, using the method described in \cite{Baluev} and as implemented in astropy \citep{astropy:2013,astropy:2018}. 

The sources that had a FAP of less than $10^{-8}$ were visually inspected, and we selected the ones that lay to the left of the main sequence in the HACKS $R_{625}$ vs $B_{435} - R_{625}$ CMD, to search for systems with a hot component, as suggested by their blue colors. Many sources had peaks in the periodogram coincident with the 96-minute HST satellite orbital period and many had a peak that was driven by outliers in their light curves, rather than periodic signals. Nonetheless, with this method, we were able to find variable sources (periodic and aperiodic ones). 

We manually vetted the variable stars reported in \citet{cohn_identification_2010,kaluzny_photometric_2006,kaluzny_time_2003}. These include the known X-ray sources, like cataclysmic variable candidates, and other known variables. 

\begin{figure}
\includegraphics[width=1.0\columnwidth]{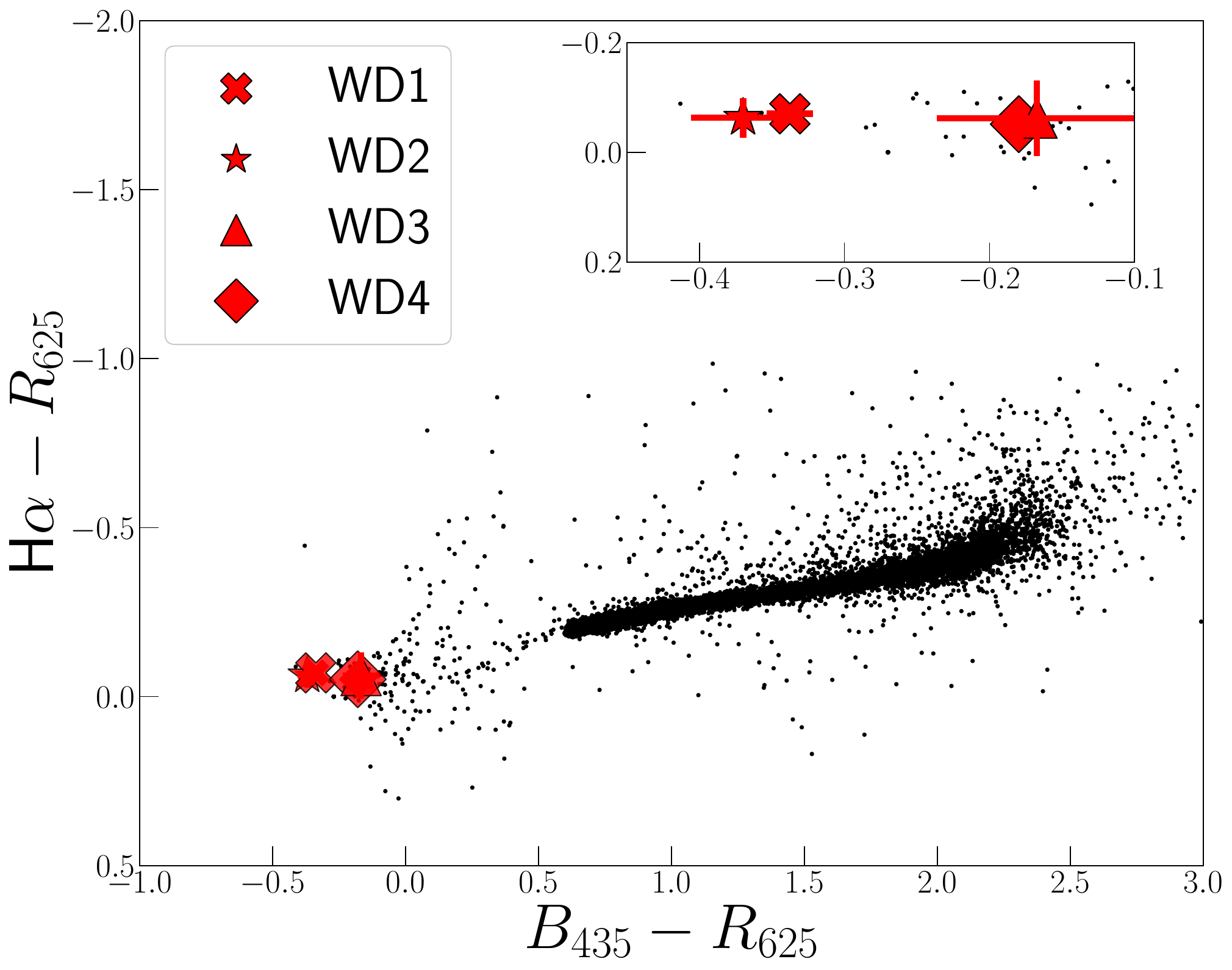}
\caption{HACKS color-color diagram for the central regions of NGC~6397. None of the sources show evidence for  $H\alpha$ emission or greater-than-average absorption. The smaller inset is a zoom-in around the colors of the variables}
\label{fig:ColorColorAll}
\end{figure}

The sources that passed the visual inspection are plotted in Fig.~\ref{fig:PanelAll}. The left part of Fig.~\ref{fig:PanelAll} shows the sources, to the left of the main sequence and near the CO WD sequence. The right panel shows another CMD, R$_{625}$ vs H$_\alpha$ - R$_{624}$, and shows 
that these sources lie 
to the right of the main sequence. We also constructed a color-color diagram, contrasting its $H\alpha$-$R$ and $B$-$R$ colors. Figure \ref{fig:ColorColorAll} shows that in such a diagram, all the sources do not show evidence for  $H\alpha$ emission, ruling out an active mass-accretion scenario, or greater-than-average absorption. Blue sources ($B_{435} - R_{625} < 0.5$) with strong $H\alpha$ excess would be on the top part of the diagram with $H\alpha - R_{625}$ less than -0.5. Blue sources with strong absorption would have $H\alpha -R_{625}$ less than 0.

The photometry for Fig.~\ref{fig:PanelAll} is from the HACKS catalogue \citep{HACKSLibralato2022}. For membership probability, we used the HUGS catalogue. The membership probabilities are reported for two of these four 
sources, but for the other two, in the catalogue they were flagged with a value of -1, indicating unavailable membership probability data. For these two sources, we relied on proper motions reported by the  HACKS catalogue. All the proper motions for all 4 reported sources are shown in Fig.~\ref{fig:PMall}, and the membership probabilities when available, along with their HACKS proper motion, are listed in Table~\ref{tab:table-detection}. The finding charts for all the sources reported are shown in the Appendix in Fig.~\ref{fig:findingcharts}.

\subsection{Variability Analysis}

For the variable sources found, we also investigated several measures of variability. 
Besides Lomb-Scargle, we searched for periodicities using the phase dispersion minimization technique \citep{PDMStellingwerf78}. To compare the degree of variability between the sources we also computed the change in magnitude in the F336W filter after getting rid of outliers (outliers were defined as points two standard deviations outside of the mean). Lastly, we computed the fractional root-mean-squared (rms) variability amplitude \citep[$F_{var}$:][]{Edelsonfvar1990ApJ...359...86E}.

\begin{equation}\label{eq:fvar}
 F_{var}  =  \sqrt{ \frac{S^2 - \overline{\sigma^2_{err}} }{\overline{x}^2}}
\end{equation}

where $S^2 $ is the observed sample variance for a time-series of 
N data points $x_i$,
and mean $\overline x$

\begin{equation}\label{eq:rms}
 S^2  =  \frac {1}{N-1}  \sum _ {i=1}^ {n}  (x_ {i}-\overline {x})^ {2} 
\end{equation}

 and  $\overline{\sigma^2_{err}}$ is the mean square error

 \begin{equation}\label{eq:meanerror}
 \overline{\sigma^2_{err}}   =  \frac {1}{N}  \sum _ {i=1}^ {n}  \sigma^2_{err,i}
\end{equation}

To calculate $\overline{\sigma^2_{err}}$, we used the errors reported by DOLPHOT. The calculated $F_{var}$, after removing the outliers, for each object is reported in Table~\ref{tab:table-detection}. We only used $F_{var}$ and the amplitude of the light curves ($\Delta U_{336}$) as metrics to compare and assess the variability between the objects found, but we did not use these metrics to find these variable sources.

\subsection{Cooling Tracks}

To estimate the mass of the candidates He WD we used the theoretical evolutionary sequences, calculated for the metallicity of NGC~6397 ([Fe/H] = -2.03),  published by \cite{Althaus2013ELMs} and shown in fig.~\ref{fig:PanelELMSTrackAll}. The authors used a statistical approach to create a grid of masses and ages for extremely low-mass He WDs. Such an approach is needed due to the fact that the different cooling tracks cross each other, and this leads to no unique solution for mass and age for a given $\log g$ and $T_{eff}$. To fit the available photometry for our targets, we used model atmospheres described at length in \cite{Rohrmann2012A&A...546A.119R} and references therein. This model atmosphere code was also used to derive the outer boundary conditions for the mentioned evolving models. Synthetic magnitudes in the HST photometry are calculated using the zero-points derived from the Vega spectrum integrated over the passband for each filter. The input parameters of model atmospheres are the effective temperature, the surface gravity (evaluated from the stellar mass and radius of the evolutionary models), and the chemical composition, which we assume here to be pure hydrogen. The gray-shaded region in fig.~\ref{fig:PanelELMSTrackAll} represents the theoretical evolutionary sequence for a He WD with $M = 0.18 M\odot$. The purple dashed line represents the theoretical cooling sequence for a He WD with $M = 0.4352 M\odot$ resulting from stable mass transfer (MT) \citep{Althaus2013ELMs}. The solid blue line is the theoretical cooling sequence for a He WD resulting from mass loss in a common envelope (CE) episode with $M = 0.4352 M\odot$. The orange dash-dotted line is the theoretical cooling curve for a CO WD from a $0.80 M\odot$ progenitor in the Zero Age Main Sequence (ZAMS) and metallicity Z=0.0001 with a $M=0.51976 M_\odot$\citep{Althaus2015A&A...576A...9A}. The black points in fig.~\ref{fig:PanelELMSTrackAll} are the dereddened HUGS catalog sources around NGC~6397. To deredden we used the extinction values from \cite{Richer2008AJ....135.2141R} and using the reddening law of \cite{Seaton1979Extinction} we get $A_{F336W} = 0.9$ and $A_{275W} =  1.11$. We assumed this extinction for all stars. 

To clean the CMD (fig.~\ref{fig:PanelELMSTrackAll}) and get rid of background galaxies and cosmic rays, we selected 'good' star candidates based on the sharp parameter, RADXS. \citep{Bedin2008,Nardiello2018MNRAS}. For the sharp parameter, we selected all the sources that satisfy the condition: $-0.15 < \text{RADXS} < +0.15$. To avoid obvious non-members of the cluster, we also only selected the sources in the HUGS catalog with a membership probability higher than 90\% or with no membership probability listed.

For fig.~\ref{fig:PanelELMSTrackAll}, the error on the F336W magnitude was calculated as the mean of the errors in their light curves and the error on the color is the error on the square root of the F275W and F336W errors added in quadrature. All the sources reported here are variable so we expect that their position on the CMD is phase-dependent and could introduce some bias in the derivation of the exact physical parameters. This is also true for the value of the adopted reddening. This is particularly relevant for the 3 WDs near the CO WD sequence. For the rest of the paper we consider all sources He WD candidates but further spectroscopic confirmation is needed.

\begin{figure}
\includegraphics[width=1.0\columnwidth]{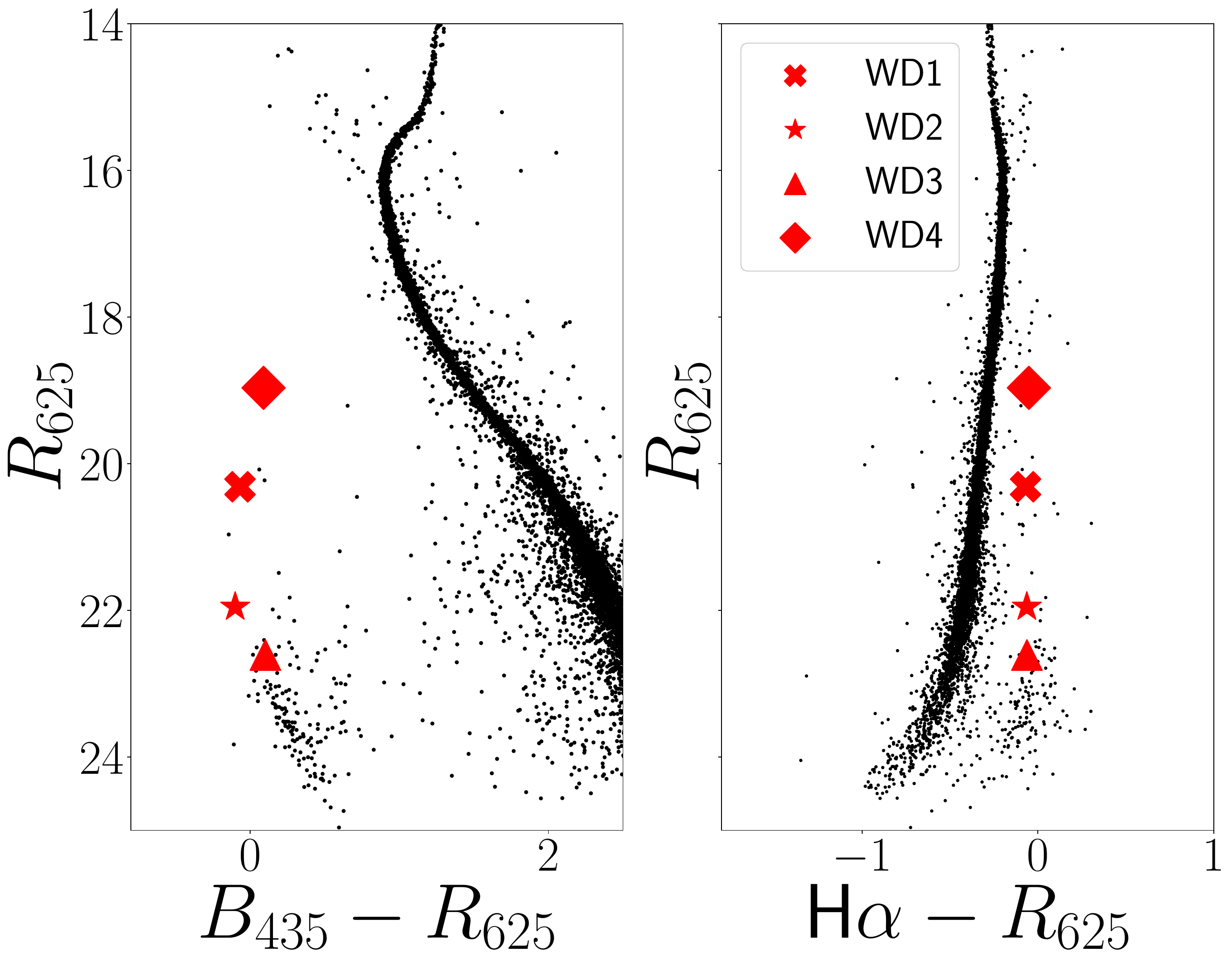}
\caption{Left: color-magnitude diagrams (CMDs) for the central regions of NGC~6397. Left: R vs B-R CMD, the red symbols show the variable 'blue' objects reported in this document. These are located to the left of the main sequence. Right: R vs H$\alpha$-R CMD. The reported sources are to the right of the MS, near the CO WD sequence. Data from the HACKS catalogue \protect{\citep{HACKSLibralato2022}.}}
\label{fig:PanelAll}
\end{figure}

\begin{figure}
\includegraphics[width=1.0\columnwidth]{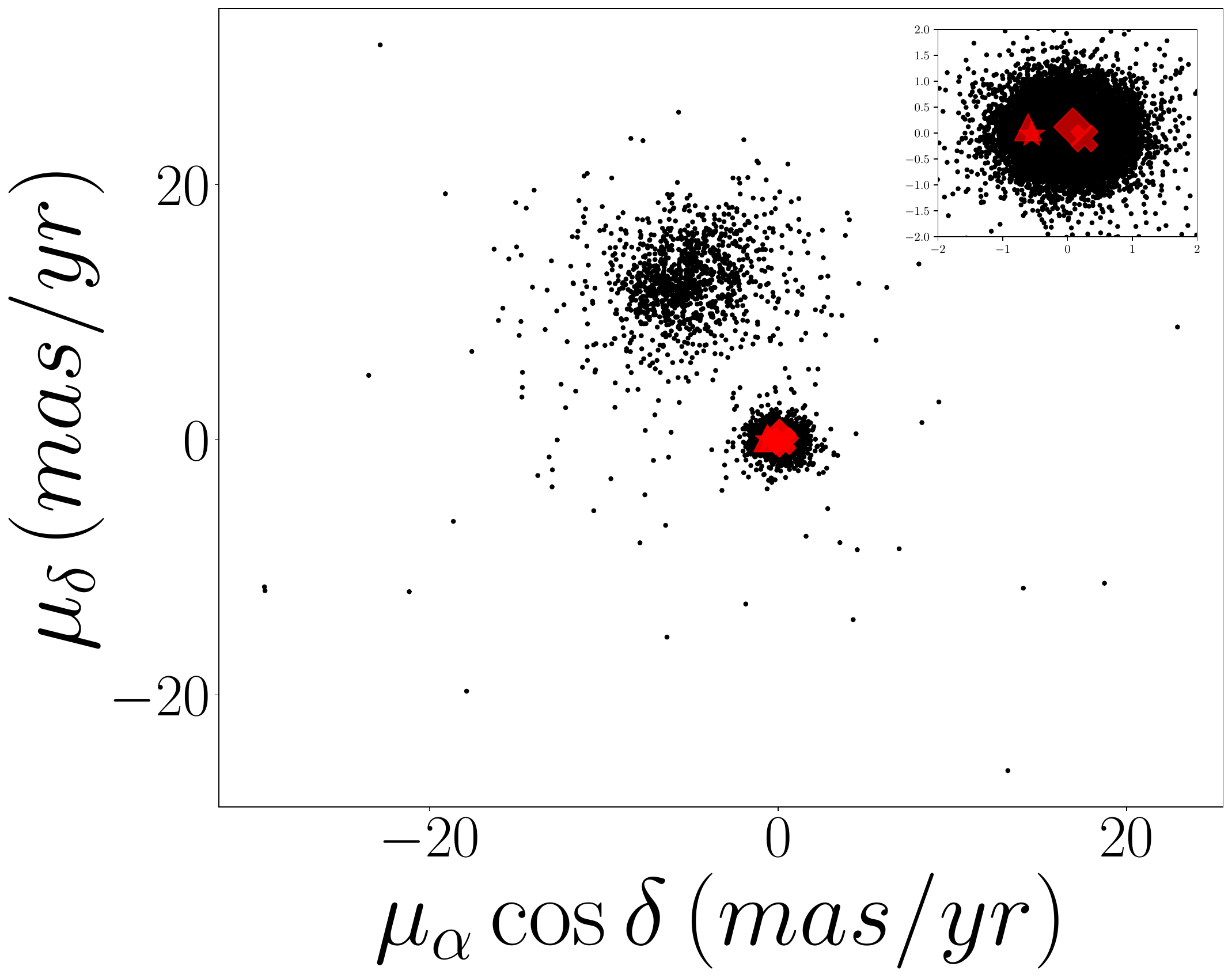}
\caption{The black points are the proper-motion components for those stars in the HACKS catalogue around NGC~6397. The red symbols are the same as the legend in fig.~\ref{fig:PanelAll} and represent all the variable WDs reported in this work. All the red symbols fall in a tight clump about the origin and are consistent with being members of the cluster. The more spread-out clump farther from the origin coordinates are field stars and not cluster members. The smaller inset is a zoom-in around the cluster members.}
\label{fig:PMall}
\end{figure}

\begin{figure}
\includegraphics[width=1.0\columnwidth]{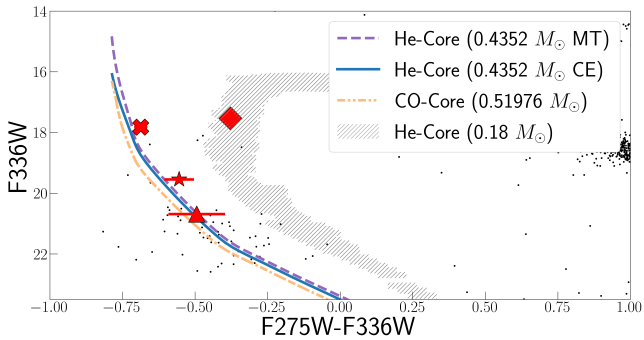}
\caption{Dereddened CMD of sources from the HUGS catalog of NGC 6397. The black small circles represent stars in the clusters, and the red symbols indicate the position of the reported variable candidate He WDs. The error on the F336W magnitude (vertical axis) was calculated as the mean of the errors in their light curves and the error on the color (horizontal axis) is the error on the square root of the F275W and F336W errors added in quadrature. The symbols are the same as the legend in Fig.~\ref{fig:PanelAll}. The gray-shaded region represents the sequence for a He WD with $M = 0.18 M\odot$\protect\citep{Althaus2013ELMs}. The purple dashed line represents the theoretical cooling sequence for a He WD with $M = 0.4352 M\odot$ resulting from stable mass transfer (MT) \protect\citep{Althaus2013ELMs}. The solid blue line is the theoretical cooling sequence for a He WD resulting from mass loss in a common envelope (CE) episode with $M = 0.4352 M\odot$. The orange dash-dotted line is the theoretical cooling curve for a CO WD from a $0.80 M\odot$ progenitor in the Zero Age Main Sequence (ZAMS) and metallicity Z=0.0001 with a $M=0.51976 M_\odot$ \citep{Althaus2015A&A...576A...9A} }
\label{fig:PanelELMSTrackAll}
\end{figure}


\section{Results and Discussion}

We present the results for each source along with a discussion regarding the possible nature of their variable behaviour. A summary of some of the properties of the sources is shown in Table~\ref{tab:table-detection}. We start with the only periodic source (WD 1) found and continue with the aperiodic variable sources.

\setlength{\tabcolsep}{3pt}
\tabletypesize{\footnotesize}
\begin{deluxetable*}{lccccccccccc}
\tablecaption{Variable Candidate He WDs in NGC 6397\label{tab:table-detection}}
\tablewidth{0pt}
\tablehead{
\colhead{ID} & \colhead{RA, Dec (J2000)\tablenotemark{a}} & \colhead{r\tablenotemark{b}} & 
\colhead{Other IDs} & \colhead{Period} & \colhead{$\Delta U_{336}$\tablenotemark{c}} & 
\colhead{$F_{var}$\tablenotemark{d}} & \colhead{Mean} & 
\colhead{$\mu_\alpha \cos \delta$\tablenotemark{e}} & \colhead{$\mu_\delta$\tablenotemark{e}} & 
\colhead{P\tablenotemark{f}} \\
\colhead{} & \colhead{} & \colhead{(\arcsec)} & 
\colhead{} & \colhead{(hrs)} & \colhead{(mag)} & 
\colhead{(mag)} & \colhead{(mag)} & 
\colhead{(mas yr$^{-1}$)} & \colhead{(mas yr$^{-1}$)} & 
\colhead{(\%)}
}
\decimals
\startdata
WD 1 & 17:40:38.1389 -53:40:35.627 & 36.0 & HUGS: 8124, HACKS: 9433 & $5.21 \pm 0.02$ & 0.11 & 0.045 & 18.76 & $0.26 \pm 0.07$ & $-0.10 \pm 0.1$  & 96.9 \\
WD 2 & 17:40:37.6872 -53:39:49.442 & 54.9 & HUGS: 12991, HACKS: 6059 & \nodata & 0.24 & 0.027 & 20.5 & $-0.55 \pm 0.1$ & $-0.03 \pm 0.11$  & \nodata \\
WD 3 & 17:40:39.2410 -53:40:33.094 & 25.9 & HUGS: 8772, HACKS: 11783 & \nodata & 0.59 & 0.098 & 21.4 & $-0.61 \pm 0.15 $ & $0.11 \pm 0.15$  & \nodata \\
WD 4 & 17:40:40.6728 -53:40:20.262 & 14.8 & NF1$^{g}$, HUGS: 10386, HACKS: 9772 & \nodata & 0.13 & 0.019 & 18.3 & $0.08 \pm 0.08$ & $0.12 \pm 0.12$ & 96.9 \\
\enddata
\tablenotetext{a}{Coordinates from the HUGS catalogue \citep{Piotto2015AJ,Nardiello2018MNRAS}.}
\vspace{-.2cm}
\tablenotetext{b}{Separation from cluster center (17:40:42.10, -53:40:27.96).}
\vspace{-.2cm}
\tablenotetext{c}{Amplitude in F336W after removing outliers.}
\vspace{-.2cm}
\tablenotetext{d}{$F_{var}$ is defined in Eq.~\ref{eq:fvar}.}
\vspace{-.2cm}
\tablenotetext{e}{Proper motions from HACKS catalogue \citep{HACKSLibralato2022}.}
\vspace{-.2cm}
\tablenotetext{f}{Membership probability from HUGS catalogue when available.}
\vspace{-.2cm}
\tablenotetext{g}{Identification Number from \citep{Cool98}}
\end{deluxetable*}

\subsection{WD 1}

WD 1 is shown as a red cross in Figs.~\ref{fig:PanelAll} and \ref{fig:PanelELMSTrackAll}. This source is in the boundary of theoretical high mass He WDs (${\sim0.32-0.46} M_\odot$) and low mass CO WDs ($M\sim 0.52 M_\odot$). This is a variable object and we expect that the position on the CMD is phase-dependent and could introduce some bias in the derivation of the exact physical parameters. For the rest of the paper, we assume it is a candidate He WD since it is closer to the theoretical cooling tracks of a He WD (see Fig.~\ref{fig:PanelELMSTrackAll}). The HUGS catalogues report a 96.9\% probability of membership.

\begin{figure}
\includegraphics[width=1.0\columnwidth]{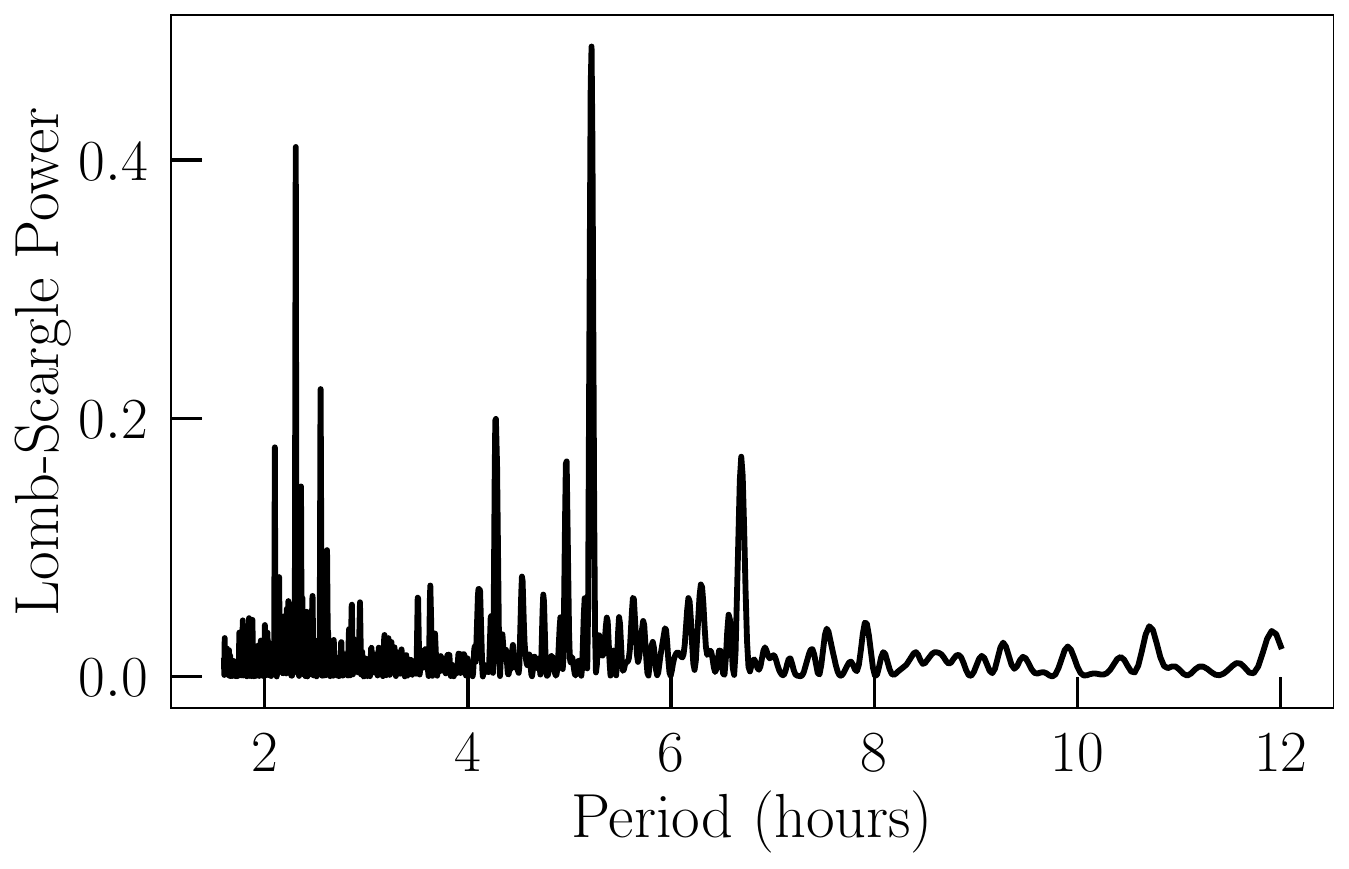}
\caption{Lomb-Scargle periodogram for WD 1. The periodogram shows an isolated peak at 5.21 hours (0.22 days). The false-alarm probability of the peak is calculated to be $3 \times 10^{-15}$, 
using the method described in {\protect\cite{Baluev}}. The amplitude of the best-fit sinusoid 0.11 magnitudes (see fig.~\ref{fig:folded}). }
\label{fig:LS}
\end{figure}

\begin{figure}
\includegraphics[width=1.0\columnwidth]{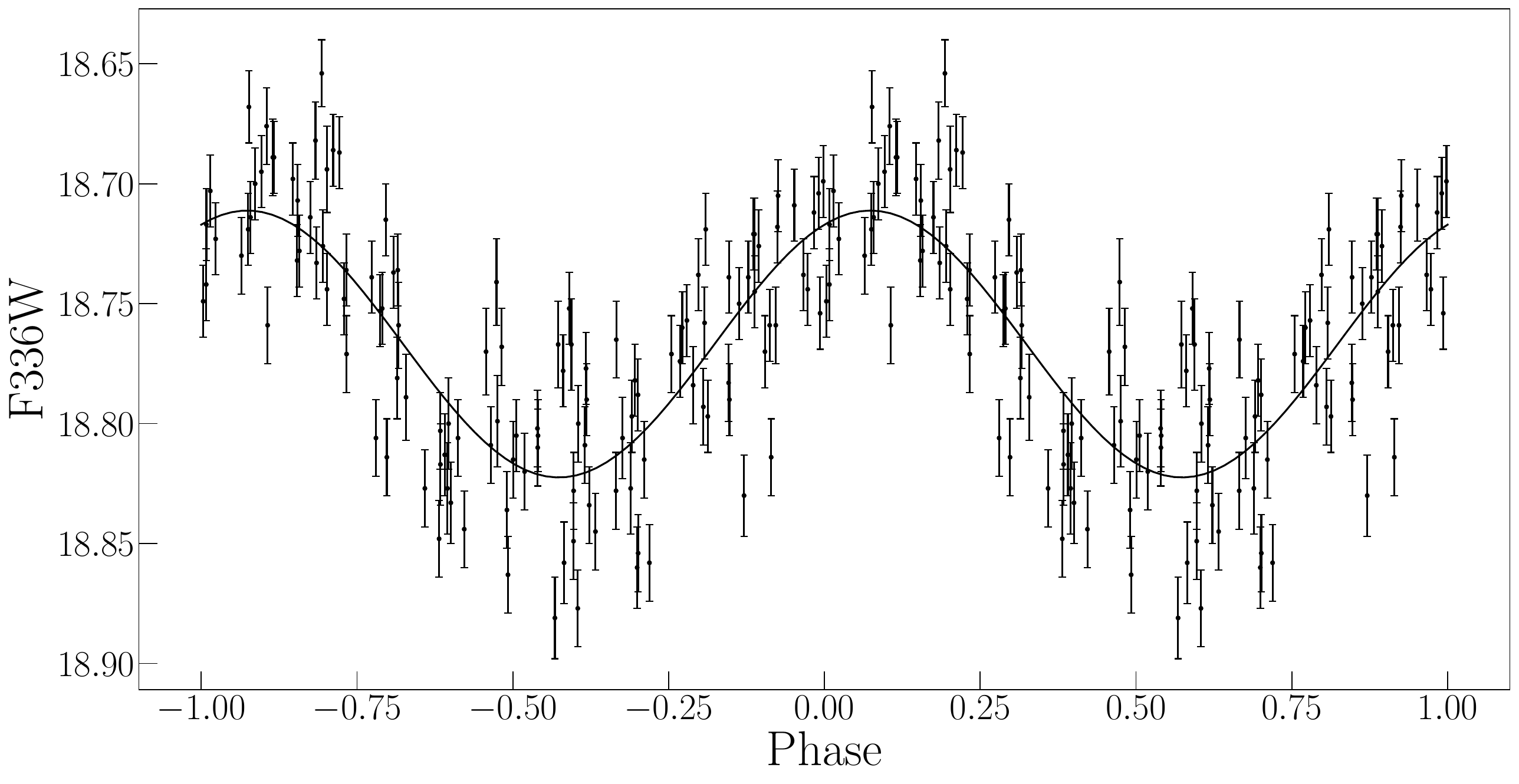}
\caption{Light curve (F336W filter) of WD 1, folded at a period of 5.21 hours or 0.22 days. }
\label{fig:folded}
\end{figure}

We performed a periodicity search on the light curve of WD 1 using the Lomb-Scargle method. The resulting periodogram is shown in Fig.~\ref{fig:LS}. The periodogram shows a significant peak with a false-alarm probability of  $3 \times 10 ^{-15}$ using the method described in \cite{Baluev} which allows getting an upper limit that is valid for alias-free periodograms. The peak is at $5.21 \pm 0.02$ hours. The uncertainty on the period was estimated by the standard deviation of the best-fitted Gaussian to the peak. The light curve folded at $5.21$ days is shown in Fig.~\ref{fig:folded}. Given that the data set spans over $100$ cycles of the period, and that the sampling is fairly dense, we can be confident that this is a real period and not due to red noise. The amplitude of the best-fit sinusoid is  0.11 magnitudes. 

We consider different scenarios for the nature of the periodicity of this He WD candidate.

\subsubsection{Pulsations}

One possibility is that the periodic optical modulation is due to pulsations of a He Core WD. CO-core WDs are known to pulsate due to global nonradial g (gravity)-mode pulsations \citep{ledoux1958}. For a review on pulsating white dwarfs, see \cite{Corsicoreview2019}. In the Galactic field the first extremely low-mass WD known to show pulsations was found by \cite{Hermes2012ApJ...750L..28H}, establishing a new class of pulsating white dwarf of extremely low mass that show long-period variability (1184-6235 s) consistent with non-radial g-mode pulsations \citep{Hermes2013MNRAS.436.3573H}. 

We reject this explanation for this system based on the time scales of known pulsations. The different time scales for known pulsating white dwarfs range from 100 s to up to 6000 s or 1.66 hours, and the periodicity of this system is 5.21 hours.



\subsubsection{Orbital Period}

Another natural explanation is that the modulation is due to orbital variation. In the Galactic field, the binary fraction of low-mass WDs is high  \citep{ELMBrown2010}. If the modulation were 
due to ellipsoidal modulation, the orbital period would be 10.6 hours. This period is dramatically longer than that at which a WD can fill its Roche lobe, so no measurable ellipsoidal modulation is expected.

He WDs are also found to be in binaries with millisecond pulsars (MSP). Heating of the white dwarf via a pulsar wind could give a periodic modulation due to the reflection effect. 
However, millisecond pulsars invariably produce detectable X-ray emission (from their heated surfaces, and possibly magnetospheric emission), typically in the range of $L_X$(0.5-10 keV)=$10^{30-31}$ erg/s \citep{Lee18,Zhao22}, though a few pulsars have $L_X$ as low as $10^{29}$ erg/s \citep[PSR J1400–1431,][]{Swiggum17}. 
NGC~6397 has relatively deep X-ray observations, with a total of 325 ks. Using the X-ray data available, \cite{BahramiaArashn2020ApJ...901...57B} studied the faint X-ray sources in NGC~6397 and established a lower limit on the X-ray luminosity of undetected sources of
$L_X = 1.0 \times 10^{29}$ erg/s. This is calculated based on 
a hypothetical source with five net counts, with an absorbed power-law spectrum (with the GC’s hydrogen column density, $N_H$, and a photon index of 1.7). 
Thus we argue that a millisecond pulsar would almost certainly have been detected as an X-ray source in this cluster.
An example of a He WD as the companion of an MSP and an orbital period of $\sim 10$ hours is 47 Tuc U \citep{Edmonds2001ApJ...557L..57E,RiveraSandoval2015MNRAS.453.2707R}. At this orbital period, the He WD companion is well inside its Roche lobe and despite having a bright X-ray source as a counterpart ($L_x = 2.6 \times 10^{30}$ erg $s^{-1}$), this system has only an orbital photometric amplitude of about 0.004 mag because the companion star subtends a very small angle as seen from the pulsar \citep{Grindlay2001Sci...292.2290G,Heinke2005ApJ...625..796H}.

Furthermore, if WD 1 was a He WD with a MSP which evolved as a primordial binary,
and assuming a WD mass of $\sim 0.4 M_\odot$, the mass–orbital period relations of \cite{Tauris1999A&A...350..928T} indicate the binary would have an orbital period of the order of several days, much longer than observed. However, in globular clusters a He WD and a MSP can exist at shorter orbital periods, likely due to stellar interactions. For example, there are systems with orbital periods of $\sim 5$ hours and with donors with $M\sim 0.5 M_\odot$ \citep{Deloye2008}. Though, we would also expect to observe X-rays coming from these systems.

All this leads us to conclude that the modulation we see in the light curves is probably not due to the orbital period. Future spectroscopic and photometric follow-up can 
measure the orbital period, if it is a binary. 
if an orbital period is confirmed, this would be one of the few binaries He WDs in any globular cluster with a measured period. This would be important to start populating the orbital period distribution of post-common envelope binaries in globular clusters.



\subsubsection{Magnetic Spots}

Another possibility we consider is that the periodic modulation is due to 
magnetic fields on 
the WD. WDs with a high enough magnetic field ($>10^4$ G) can show modulation in their light curves on the order of $>0.1$ mag due to surface magnetic spots on the WD, via the process of magnetic dichroism \citep{1976ApJ...209..208L,FerrarioMagneticWD}. \cite{PichardoMarcano2023} found this to be the most plausible explanation for another He WD in this cluster that showed a periodic optical modulation. If this is confirmed as a candidate magnetic He WD, it would be the second candidate in this globular cluster. Recently, \cite{2024MNRAS.528.6056H} investigated the role of rotation on magnetic field generation, including measurements made with the Transiting Exoplanet Survey Satellite (TESS) light curves, which are not biased towards shorter periods. These authors found that the magnetic WD spin periods are systematically shorter than those of non-magnetic WDs. Both sources therefore, upon confirmation, provide a new opportunity to study the different relationships related to magnetic field generation and appearance, in particular the dependence on the rotation of the origin of magnetic fields in He WDs.

For CO WDs, \cite{Hermes2017Rotation} found evidence for a relationship between high mass and fast rotation. This is for single-star evolution, but as we populate the rotational period distribution of candidate He WDs we can explore if the relationship works for these products of binary evolution.

\subsection{Sources with aperiodic variability}

\subsubsection{WD 2}

\begin{figure}
\includegraphics[width=1.0\columnwidth]{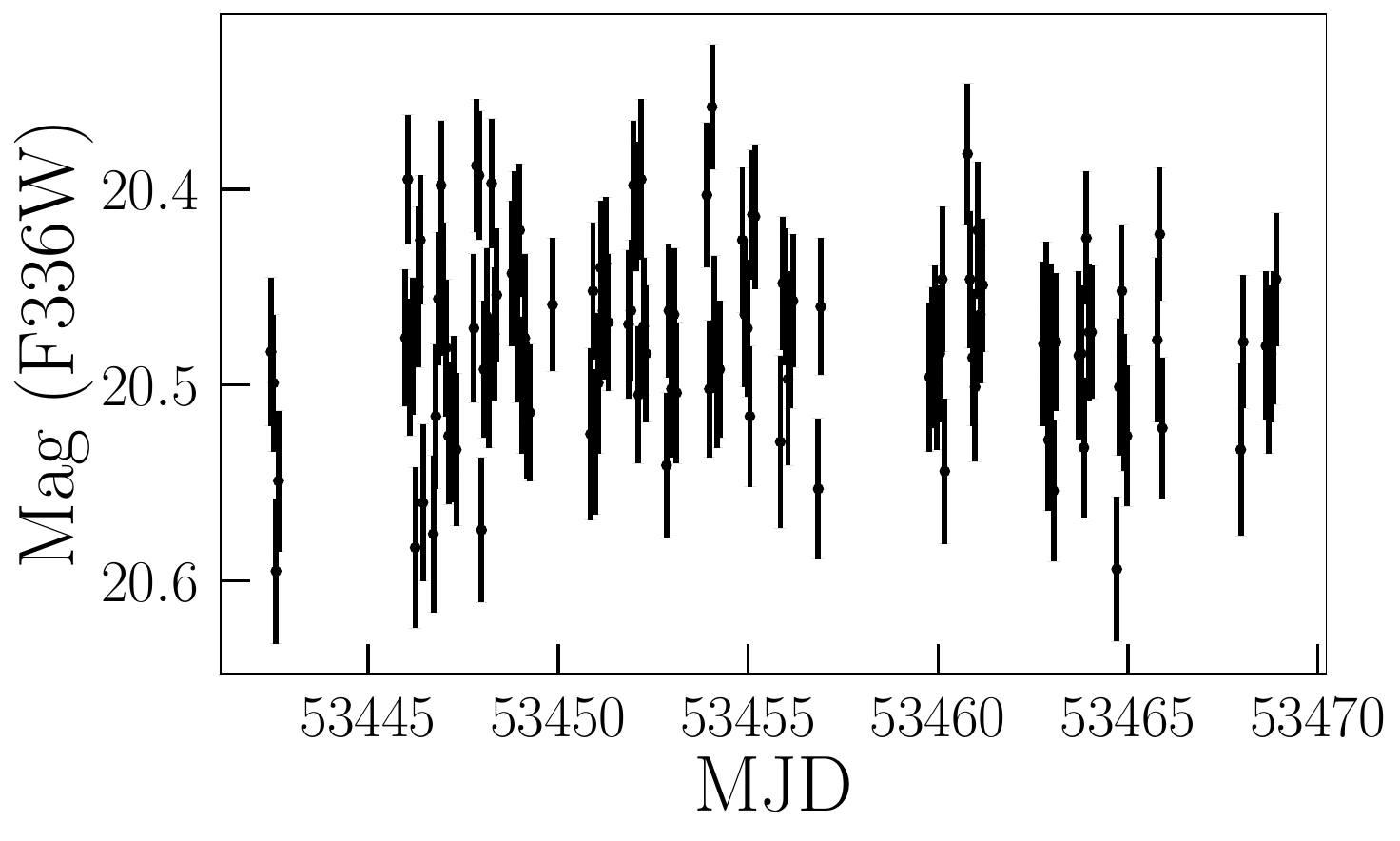}
\caption{Light curve (F336W filter) of WD 2 }
\label{fig:LC2932}
\end{figure}

WD 2 is a second variable candidate He WD. The location of WD 2 is shown by a red star in figs. \ref{fig:PanelAll} and \ref{fig:PanelELMSTrackAll}. Like WD 1, this source is in the boundary of theoretical high-mass He WDs ($M\sim 0.4 M_\odot$) and low-mass CO WDs ($M\sim 0.5 M_\odot$). We also assume it is a candidate He WD since it is closer to the theoretical cooling tracks of a He WD (see Fig.~\ref{fig:PanelELMSTrackAll}). 

No membership probability is available from the HUGS catalogue but it is a probable member based on the proper motion given by the HACKS catalogue (see Fig.~\ref{fig:PMall}). The light curve of WD 2 is shown in Fig.~\ref{fig:LC2932}. The light curve shows a change in magnitude of $~0.2$ mag. No real periodicity was found using Lomb-Scargle and Phase Dispersion Minimization. Because our sampling is largely on the 96-minute satellite orbital period, faster periodic variations can be hard to detect, even if the variability from them is real. The same X-ray limit of $L_X < 9.76 \times 10^{28}$ erg/s  established by \cite{BahramiaArashn2020ApJ...901...57B} applies. Spectroscopic follow-up can help determine the nature of this object and the reason for the variability. This is one of the least variable sources from all the reported sources. The variance, $S^2$, is close to the mean square error, $\overline{\sigma^2_{err}}$, ($S^2/\overline{\sigma_{err}^2} = 1.6$). This gives a $F_{var} = 0.027$.



\subsubsection{WD 3}

\begin{figure}
\includegraphics[width=1.0\columnwidth]{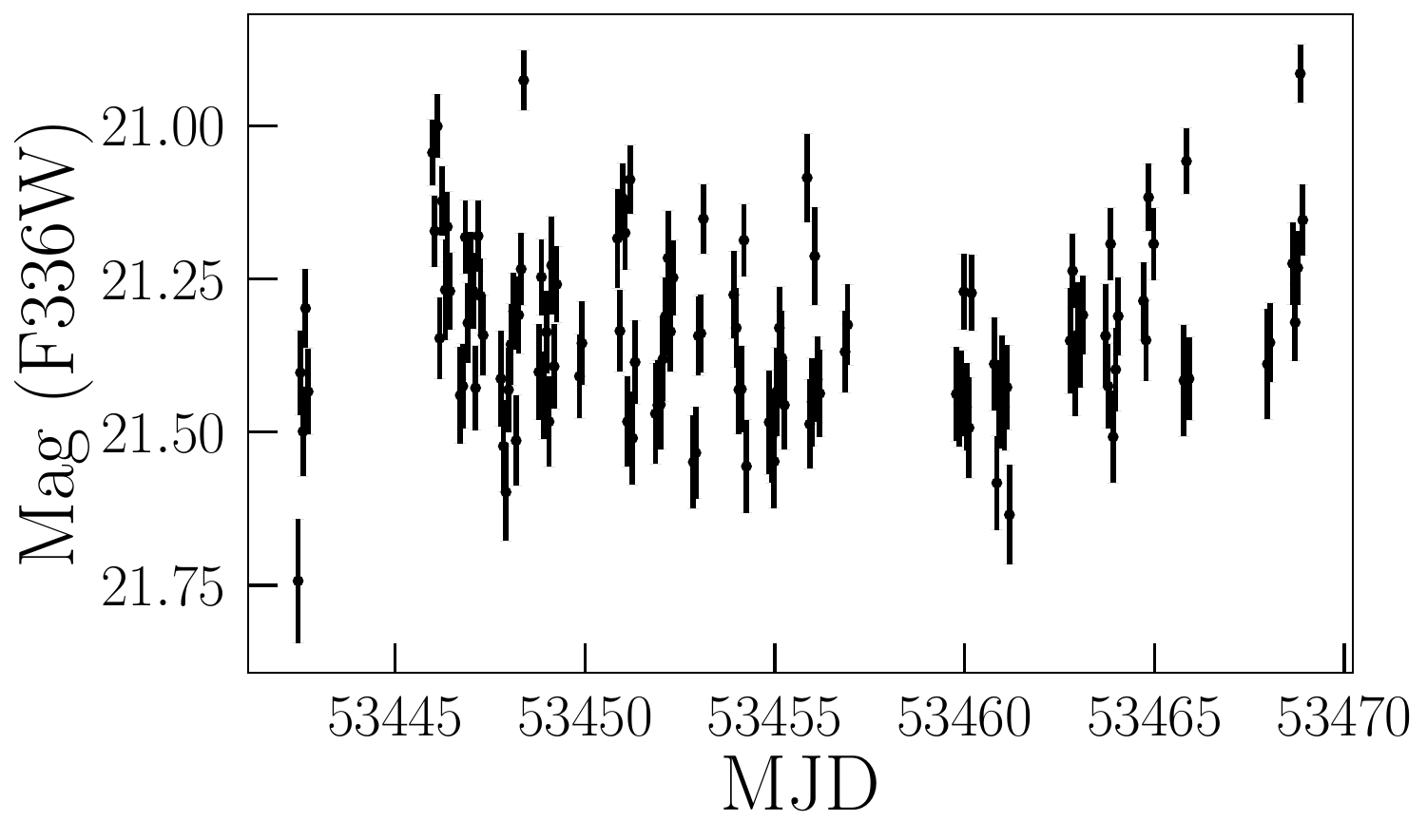}
\caption{Light curve (F336W filter) of WD 3}
\label{fig:LC3399}
\end{figure}

WD 3 is shown as an upright red triangle in figs.~\ref{fig:PanelAll} and \ref{fig:PanelELMSTrackAll}. It is the closest to the CO WD sequence with a mass of $\sim 0.52 M_\odot$, and its position on the CMD is consistent with being a CO WD or a He WD (see Fig.~\ref{fig:PanelELMSTrackAll}).  No membership probability is available from the HUGS catalogue but it is a probable member based on the proper motion given by the HACKS catalogue (see Fig.~\ref{fig:PMall}). This WD candidate shows the largest photometric variability of all the candidates shown here. The light curve is shown in Fig.~\ref{fig:LC3399} and shows a change in magnitude of 0.75 mag. After removing the outliers, as defined to be two standard deviations outside the mean, the variability is still significant with a change in magnitude of $\sim 0.6 $. No periodicity was found but it is a good source for spectroscopic follow-up to determine the physical mechanism for a highly variable WD in NGC~6397. This is the most variable source reported here with an $F_{var} = 0.098$.

\subsubsection{WD 4}


The last variable candidate WD we found is WD 4. WD 4 is shown as a red diamond in Figs.~\ref{fig:PanelAll} and \ref{fig:PanelELMSTrackAll}.  This is the variable He WD candidate with the lowest estimated mass ($\sim 0.18 M_\odot$ based on the position in the CMD, see Fig.~\ref{fig:PanelELMSTrackAll}). It is a probable cluster member with a $96.9\%$ membership probability reported in the HUGS catalogue. This source was first found by \cite{Cool98} as a non-flickering UV star (NF1). It was later spectroscopically confirmed as a He WD by \cite{edmonds_cataclysmic_1999}. They reported a best fit for their models of a He WD with   $T_{eff} = 5 000 K $, $\log g = 6.25 \pm 1.0 $, and a mass in the range of $0.2-0.5 M_\odot$. From the $H_\beta$ line they measured a redshift of $247 \pm 50$ km/s, which they attributed to an unseen binary companion. This source is also near the center of the cluster ( $14.8^{\prime \prime}$), 
likely
due to mass segregation in the cluster. 

\begin{figure}
\includegraphics[width=1.\columnwidth]{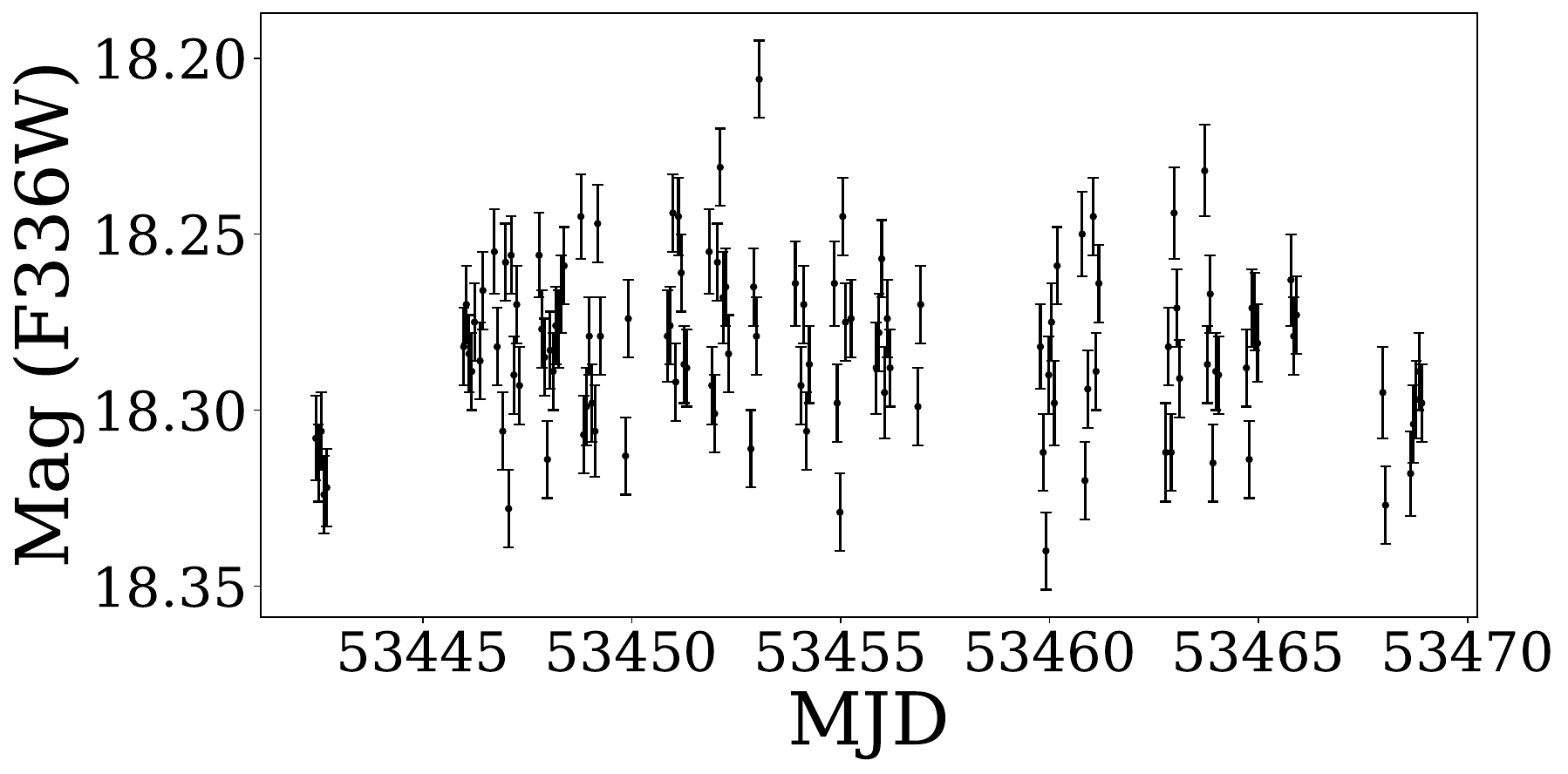}
\caption{Light curve (F336W filter) of WD 4. }
\label{fig:LCWD4}
\end{figure}

Using observations from the Multi Unit Spectroscopic Explorer (MUSE) \cite{gottgens_2023_8355370} found radial velocity variations in the spectra and reported a periodicity of $0.54$ days, which is probably this system orbital period. We searched for a periodicity in the light curve (Fig.~\ref{fig:LCWD4}) but were not able to find a significant peak in the periodogram.


\section{Conclusion}

In this paper, we reported the discovery of 4 variable WD candidates in NGC~6397. 

For WD~1 we found a periodicity of $5.21 \pm 0.02$ hours. Lacking other data besides color (see Fig.~\ref{fig:PanelELMSTrackAll}), it is hard to confirm the true cause of the periodic optical modulation in the light curve. We consider three possible scenarios: pulsations, orbital variability, and rotation of a magnetic spot. Based on the time scales we argue against pulsations. Based on the lack of X-rays, 
this periodicity is hard to explain as orbital variation caused by a pulsar companion. We argue that the periodicity in this system can be interpreted as a periodic modulation due to a magnetic spot on a rotating candidate He WD. This makes WD~1, the second candidate magnetic WD in NGC~6397 \citep{PichardoMarcano2023}.

We also report 3 other WDs candidates that showed variability on the order of a few tenths of a magnitude. Using Lomb-Scargle and phase dispersion minimization techniques, we were not able to find a periodicity for these 3 sources. This could be due to the periodicity being shorter than the Nyquist frequency, or that the nature of the variability is aperiodic. The object with the largest change in magnitude was WD 3 with a change in magnitude of $~0.6$ mag. 

All these sources are good spectroscopic follow-up targets to explain the observed variability. Confirming the binarity and the orbital period of possible double degenerate binaries, like He WDs are thought to be, and comparing it to the galactic field would give us a test on binary evolution models in dense environments and how much dynamics and metallicity change their formation and evolution. The variability of these objects could be attributed also to rotation or pulsations of an isolated or binary He WD. Follow-up spectroscopy can help find evidence of magnetism in these systems by, for example, measuring the Zeeman effect on their spectra.

Overall this work shows the benefits of high cadence photometric surveys in globular clusters to find variable WDs. 



\section*{Acknowledgments}

We thank the anonymous referee for a constructive
report which has improved the clarity and quality of this paper. This work used public data available in the Mikulski Archive for Space Telescopes under the programs GO-10424, GO-10257 and GO-13297. The specific observations analyzed can be accessed via \dataset[doi:10.17909/A1RR-GF41]{https://doi.org/10.17909/A1RR-GF41}. COH is supported by NSERC Discovery Grant RGPIN-2023-04264. DB acknowledges partial support from FONDECYT through grant 3220167. MPM acknowledges support from the EMIT NSF grant (NSF NRT-2125764). MPM would like to thank Ilija Medan for his help with \latex{} and Sam Obenchain for his support and interesting discussions. This research has made use of the Science Explorer, funded by NASA under Cooperative Agreement 80NSSC21M00561.

%

\vspace{5mm}


\software{PyAstronomy \citep{pyastronomypackage},  
          SciPy \citep{SciPy-NMeth}, 
          Astropy \citep{astropy:2013,astropy:2018}, Matplotlib \citep{matplotlib} APLpy \citep{aplpy2012,aplpy2019}.
          }



\appendix
\pagebreak

\section{Finding Charts}

\begin{figure}[h]
    \centering
     \subfloat{%
        \includegraphics[width=0.4\linewidth, height=0.4\textwidth]{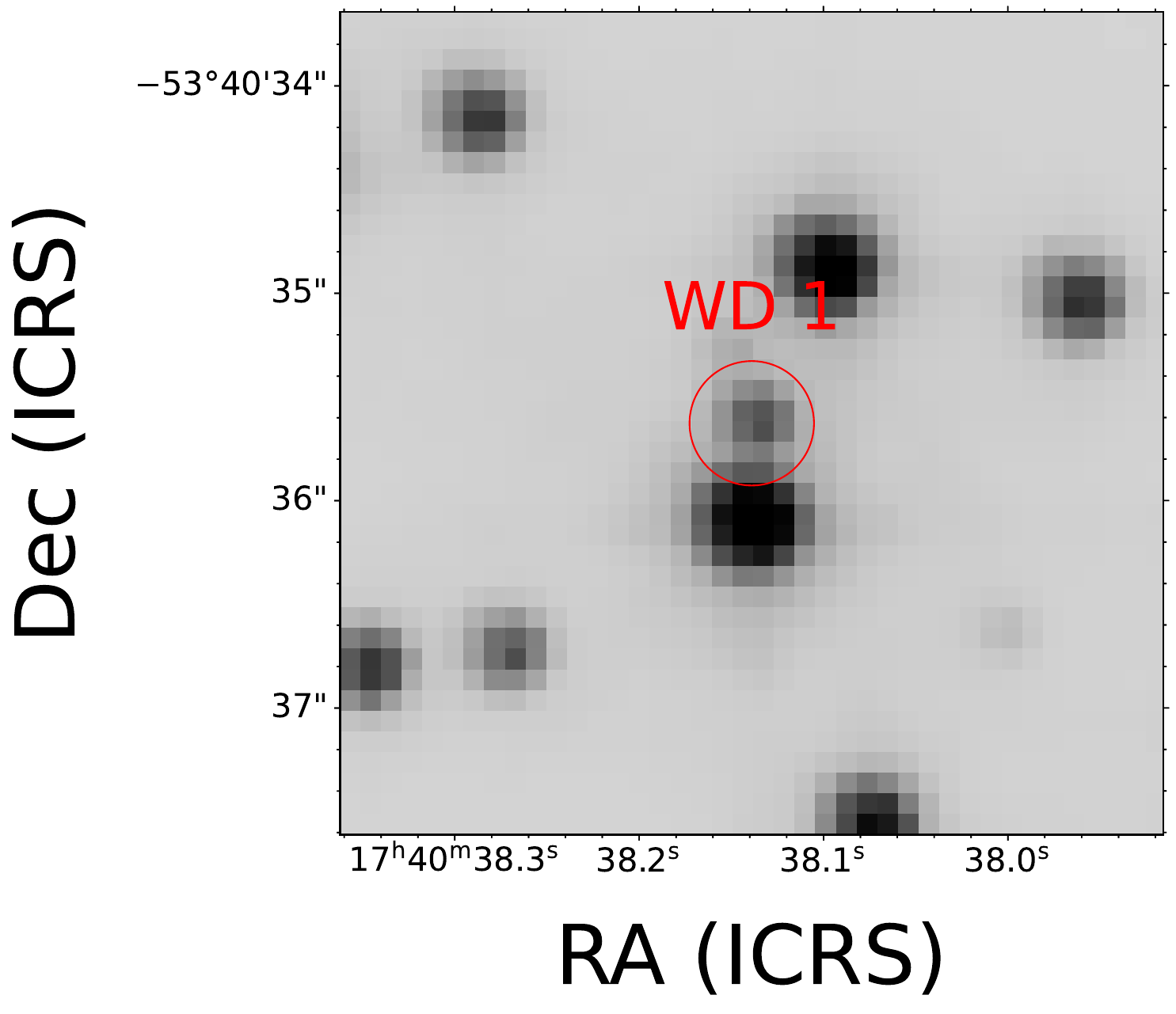}%
        \label{fig:gull}
                \includegraphics[width=0.4\linewidth, height=0.4\textwidth]{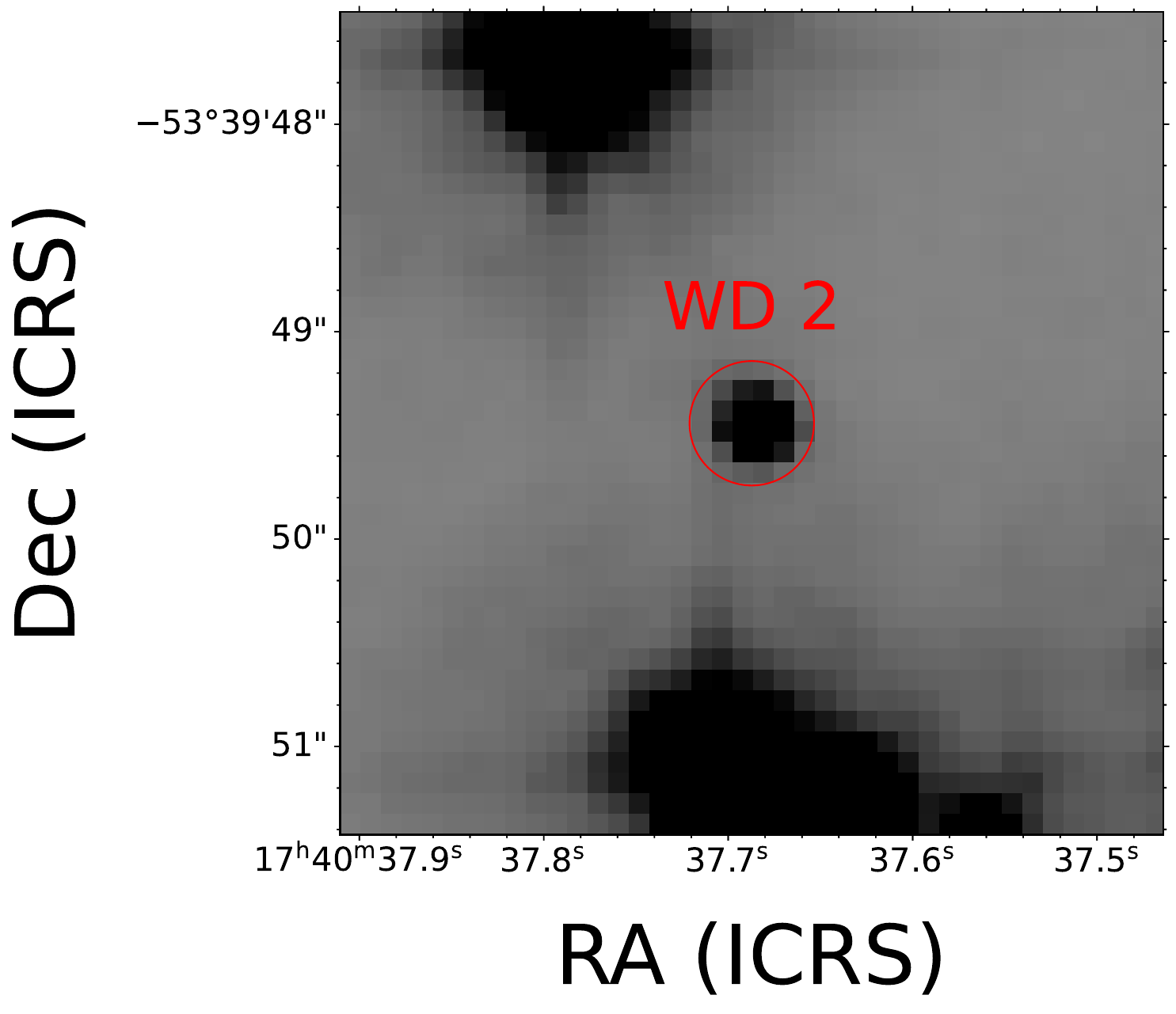}
        \label{fig:tiger}%
        }

     \subfloat{%
        \includegraphics[width=0.4\linewidth, height=0.4\textwidth]{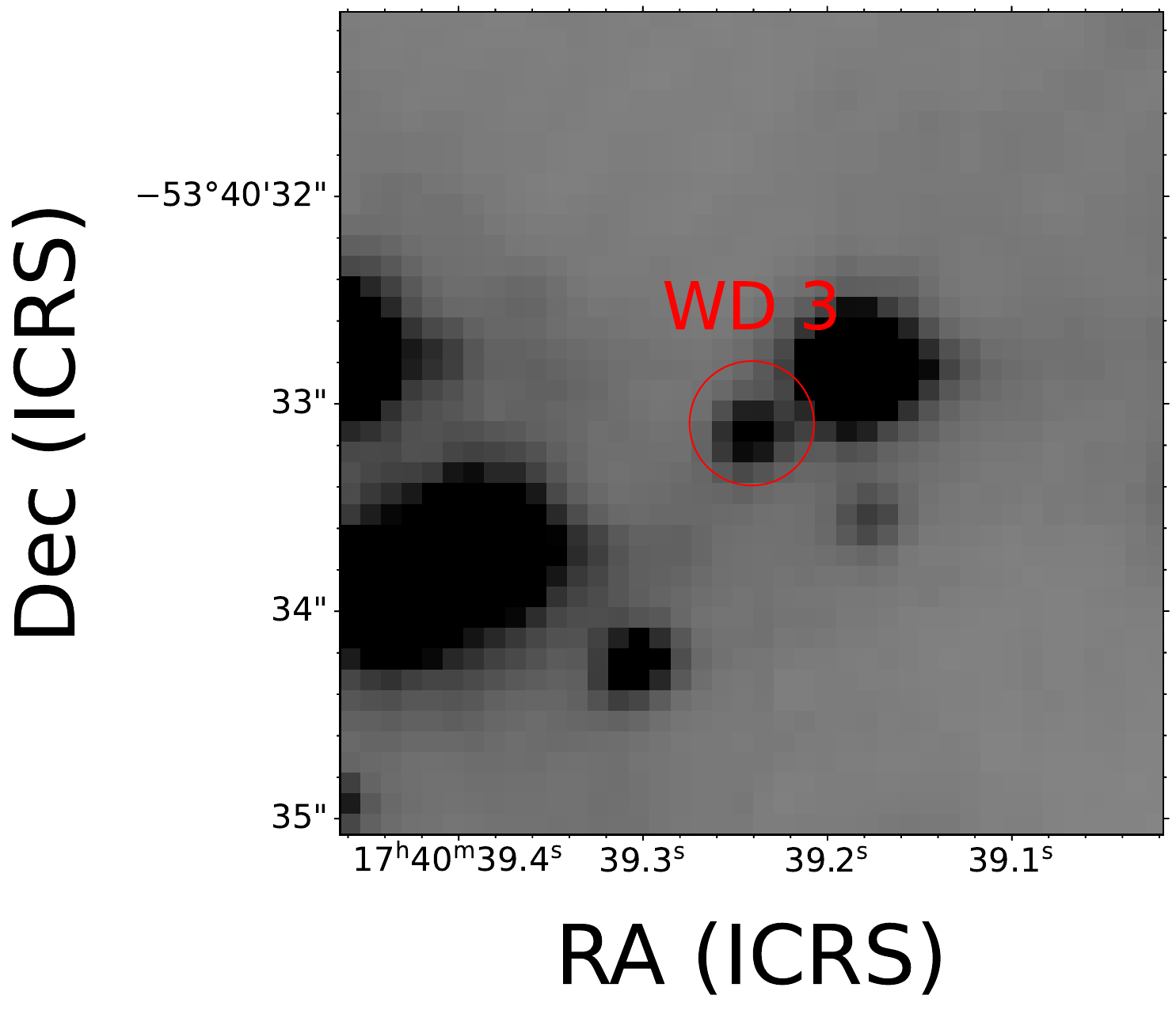}%
        \label{fig:gull}
              \includegraphics[width=0.4\linewidth, height=0.4\textwidth]{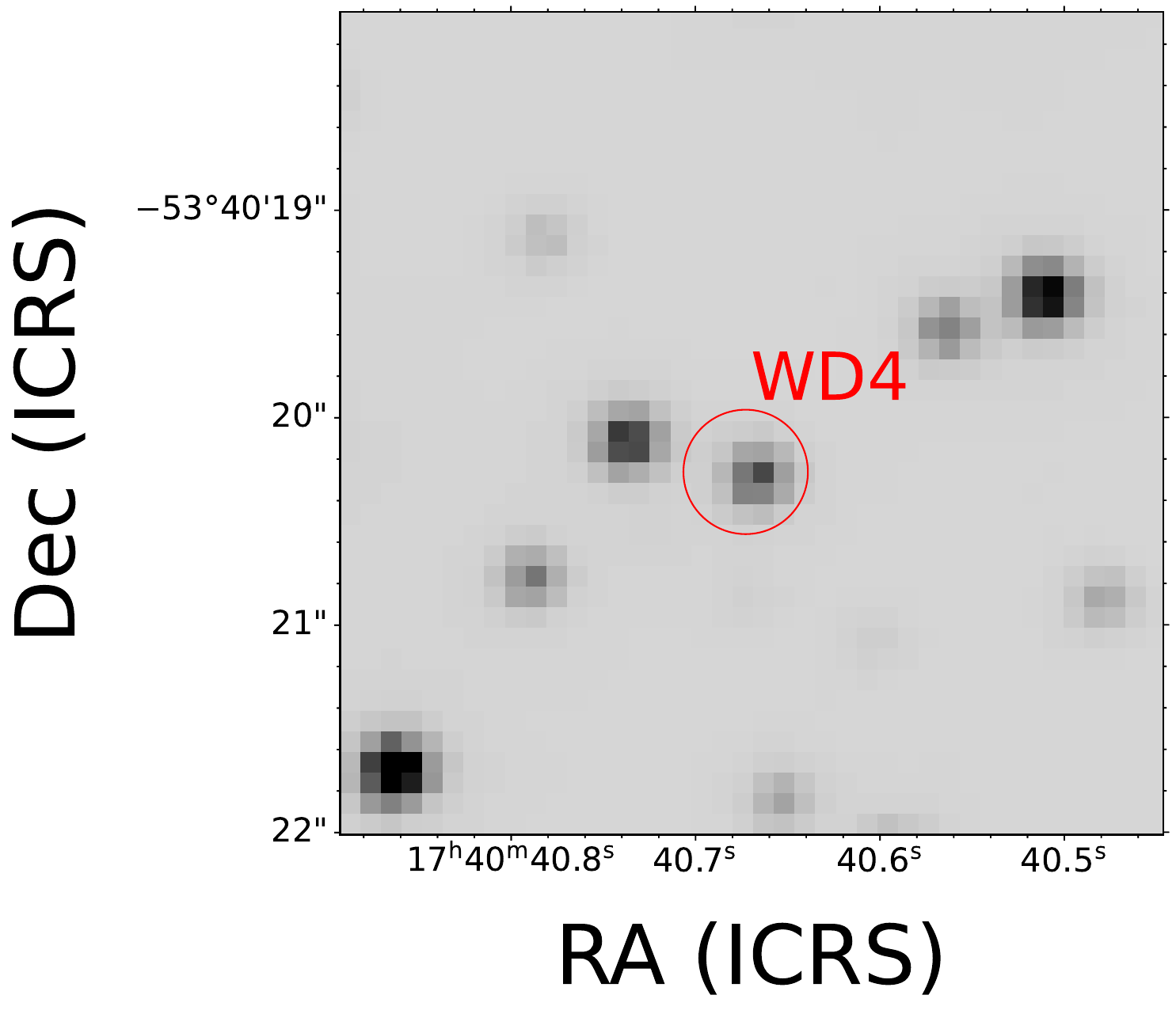}%
        \label{fig:tiger}
        }
        \label{fig:gull}
    \caption{Top: Finding chart for WD 1 (left), WD 2 (right) in the F3336W filter.
    Bottom: Finding chart for WD 3 (left), WD 4 (right) in the F3336W filter. The red circles are 0.3" in radius and centerd at the positions listed in table~\ref{tab:table-detection} for each object. In all images, north is up.}\label{fig:findingcharts}
\end{figure}


\bibliography{example}{}
\bibliographystyle{aasjournal}



\end{document}